# Hidden area and mechanical nonlinearities in freestanding graphene

Ryan J.T. Nicholl[1], Nickolay V. Lavrik[2], Ivan Vlassiouk[3], Bernadeta R. Srijanto[2], and Kirill I. Bolotin[1,4]

[1]Department of Physics and Astronomy, Vanderbilt University, Nashville, TN 37235, USA
[2]Center for Nanophase Materials Sciences, Oak Ridge National Laboratory, Oak Ridge, TN 37831, USA
[3]Energy & Transportation Science Division, Oak Ridge National Laboratory, Oak Ridge, TN 37831, USA
[4]Department of Physics, Freie Universität Berlin, Arnimallee 14, Berlin 14195, Germany

**We investigated the effect of out-of-plane crumpling on the mechanical response of graphene membranes. In our experiments, stress was applied to graphene membranes using pressurized gas while the strain state was monitored through two complementary techniques: interferometric profilometry and Raman spectroscopy. By comparing the data obtained through these two techniques, we determined the geometric hidden area which quantifies the crumpling strength. While the devices with hidden area $\sim 0\,\%$ obeyed linear mechanics with biaxial stiffness $428 \pm 10$ N/m, specimens with hidden area in the range $0.5 - 1.0\,\%$ were found to obey an anomalous nonlinear Hooke's law with an exponent $\sim 0.1$.**

A thin membrane is always crumpled due to its low bending rigidity and resulting inability to sustain compressive forces. Such crumpling has been actively investigated during the last three decades to describe the behaviors of wrinkled skin [1,2], biological lipid membranes [3,4], and solar sails [5]. The advent of graphene and other 2D materials allowed testing the models of crumpling in crystalline membranes at the ultimate atomic thickness limit [6]. In graphene specifically, crumpling originates from static wrinkling [7–9] and out-of-plane (flexural) phonons [10–12] and persists in both supported and free-standing samples [13]. Recent theoretical work showed that every mechanical property of graphene is renormalized due to crumpling [14–20]. In particular, crumpling causes the reduction of the stiffness [21,22], increased bending rigidity [11,23], variable (and negative) Poisson's ratio [24,25], and negative thermal expansion [26,27]. At the same time, the contribution due to crumpling is almost universally ignored in the experiments probing mechanics of these materials. This may lead to misinterpretation or incorrect conclusions, for example, while using graphene nanoelectromechanical (NEMS) devices to detect mass, force, or displacement. Experiments that do probe the interplay between crumpling and graphene mechanics remain highly challenging [21,23,28].

Previously, we developed an approach to probe the mechanical response of crumpled graphene membranes [29]. We observed the reduction of graphene stiffness down to ~20 N/m and hypothesized that it was mostly due to static wrinkling. Unfortunately, the electrostatic actuation scheme used in that work prevented us from applying sufficient stress to change the crumpling strength. Because of that, while the hints of nonlinear behavior in stress-strain curves were observed, we could not investigate it in detail.

The goal of this work was to study the transition of graphene membranes from the crumpled state characterized by reduced stiffness to the flat state with accepted stiffness close to $400$ N/m (Young's modulus ~1 TPa). To apply mechanical stress sufficient to drive this transition, some membranes were pressurized with compressed gas while others were pre-stressed during fabrication. To characterize the transition, we quantified the degree of crumpling by comparing the measurements of strain via Raman spectroscopy and wide-field interferometry. These experimental innovations allowed the observation of a nonlinear Hooke's law in samples with different amounts of crumpling. Our findings were confirmed by the comparison with quantitative theory.

**Experimental Setup**

Two types of samples were produced: standard and strain-engineered. Both sample types were prepared by the wet transfer of graphene grown via chemical vapor deposition (CVD) with subsequent thermal annealing as described in a previous work [29]. Standard samples consisted of a monolayer graphene membrane suspended over a single hole with diameter $\sim 10$ µm in a silicon nitride (SiN$_x$) support on a silicon chip (Fig. 1a left). To create strain-engineered samples, we patterned an additional $\sim 50 - 100$ nm deep, 5 µm wide recess in the SiN$_x$ around the edges of the hole (Fig. 1a, right). Graphene was pulled into the recess by van der Waals forces during transfer. From geometrical considerations, this process is expected to impart $\leq 1\%$ strain on graphene. Strain-engineered samples allow us to extend the range of applicable stress and act as an experimental control for flat graphene subjected to perfectly in-plane and uniform built-in stress.

The mechanical response of graphene membranes was characterized through measurements of sample deflection under a known pressure ($P$). Pressure was applied to graphene using compressed nitrogen gas [30]. The gas was fed into a cell that was sealed with the graphene membrane on one side. A PDMS O-ring established a leak-tight seal between SiN$_x$/Si sample substrate and the cell base (Fig. 1b). A digital pressure gauge and a gas flow regulator allowed the control of pressure in increments $< 1$ KPa up to $\sim 200$ KPa, $\sim 10$ times larger than in our previous work [29]. At pressures $> 200$ KPa the O-ring fails. The pressure was stable to below $0.05$ KPa over the length of our measurements, $\sim 1$ hour. Identical responses for both positive and negative pressures (colored vs. grey curves in Fig. 1c) confirm that there was no significant slippage or delamination occurring between graphene and the SiN$_x$ interface. This conclusion is also consistent with the lack of discontinuities in pressure vs. deflection curves [30,31] or consequently the strain vs. stress curves discussed later. From pressure, we determined the radial stress [32] of graphene $\sigma = Pa^2/4h$, where $h$ is center point displacement determined from interferometry described below and $a \sim 5$ µm is radius of the device. We note the stress $\sigma$ is the total stress that includes both the built-in (existing without the application of pressure) and applied (due to applied pressure) stress components. Consequently, $\sigma = 0$ means the membrane is completely relaxed.

Upon application of pressure, the mechanical strain $\varepsilon$ of the graphene membrane was measured in two different yet complementary approaches: interferometric profilometry ($\varepsilon_{Int}$) and Raman spectroscopy ($\varepsilon_{Ram}$). The strain $\varepsilon$ determined by both measurement types is applied strain. By definition, $\varepsilon = 0$ at zero applied pressure. In the first method, the deflection of graphene is probed via wide-field phase shift interferometry using 530 nm, $< 0.1$ mW power illumination. This allowed the direct determination of lateral membrane topography on the micron scale (Fig. 1c) and the measurement of the center point deflection ($h$) with nanometer resolution. From geometrical considerations, the radial strain [32] was then determined as $\varepsilon_{Int} = 2h^2/3a^2$. Since $\varepsilon_{Int}$ is measured geometrically relative to the initial state at $P = 0$, it does not include the built-in strain ($\varepsilon_0$) component.

In our second method, the strain was determined by monitoring the shifts of the 2D and G peaks in the Raman spectra of graphene taken at the center of the membrane. Inaccuracy of spot position by up to 2 μm changes the results no more than 4 %, see Supplemental Material (SM) Fig. 2 [33]. We use a focused 633 nm excitation source with an estimated spot size < 1 μm, resolution ~1 cm$^{-1}$ and power < 1 mW to avoid heating (Fig. 1d). The strain was extracted as: $\varepsilon_{Ram}^{2D,G} = (\partial\omega^{2D,G}/\partial\varepsilon_{Ram}^{2D,G})^{-1}(\omega^{2D,G} - \omega_0^{2D,G})$ [34]. Here $\omega^{2D,G}$ is the frequency position of the 2D(G) peak of strained graphene and $\omega_0^{2D,G}$ is the position of the same peak at zero applied pressure. In this way, $\varepsilon_{Ram}$ is also a measurement of strain relative to the initial state [35]. The peak sensitivity for each device was found by extracting the slope of Raman peak positions vs. $\varepsilon_{Int}$ (Fig. 2a, left inset, dashed line) at stresses > 1 N/m. We find peak sensitivities $|\partial\omega^{2D}/\partial\varepsilon_{Int}|$~155 − 200 cm$^{-1}$/% and $|\partial\omega^G/\partial\varepsilon_{Int}|$~55 − 90 cm$^{-1}$/% consistent with recent values in literature [34,36–38]. The necessity of applying such large stress is discussed later. We ensured that changes in Raman peak positions vs. pressure were entirely due to strain rather than e.g. changes in doping by observing $\partial\omega^{2D}/\partial\omega^G$~2.2 (Fig 2b right inset) [39]. This also confirms identical results for extraction of strain from either G or 2D peaks.

**Comparison of stress-strain curves from interferometry and Raman spectroscopy**

The stress-strain relationships of three standard samples (A, B, and C) as measured from Raman spectroscopy, $\varepsilon_{Ram}(\sigma)$, and interferometry, $\varepsilon_{Int}(\sigma)$, are shown in Fig. 2a, b. We observe dramatic differences between the $\varepsilon_{Ram}(\sigma)$ and $\varepsilon_{Int}(\sigma)$ curves. The $\varepsilon_{Ram}(\sigma)$ curves are linear (Fig. 2a). The average biaxial modulus for all devices extracted from them is $\tilde{E}_{2D} = d\sigma/d\varepsilon_{Ram} = 480 \pm 10$ N/m. In contrast, the $\varepsilon_{Int}(\sigma)$ curves are strongly non-linear (Fig. 2b). In the region of low stress ($\sigma < 1$ N/m), graphene is soft, $\tilde{E}_{2D}$~30 − 150 N/m. At the same time, in the high stress region ($\sigma > 1$ N/m) we retrieve an average value of $\tilde{E}_{2D} = 450 \pm 70$ N/m, close to what is measured by Raman spectroscopy. In the most interesting intermediate region ($\sigma$~1 N/m), we see a transition from non-linear to linear mechanical response with increasing stress. For the strain-engineered device (Fig. 2a, b, orange points), we observe a linear and identical response from both Raman spectroscopy ($\tilde{E}_{2D} = 430 \pm 10$ N/m) and interferometry ($\tilde{E}_{2D} = 426 \pm 7$ N/m) throughout the range of applied stress.

We note that the biaxial moduli measured from Raman spectroscopy or from interferometry at high stress are close to the values obtained in other experiments [40–42], consistent with the value for flat graphene, $\tilde{E}_{2D}$~400 N/m calculated from Lamé parameters [27] ($\lambda = 2$ eV Å$^{-2}$ and $\mu = 10$ eV Å$^{-2}$) and extracted from simulations [43]. The biaxial modulus can be converted to an in-plane stiffness, $E_{2D} = (1 − \upsilon)\tilde{E}_{2D}$ where $\upsilon$~0.165 is the commonly used value for the Poisson's ratio of graphene [44]. This yields an average of $E_{2D} = 380 \pm 30$ N/m over all our devices. This corresponds to a Young's modulus of ~1 TPa. However, the Poisson's ratio for graphene is not well known and may not be constant or even take negative values [24,25]. Therefore, we directly report the biaxial modulus $\tilde{E}_{2D}$.

The data of Fig. 2 invites the following questions. Why are the observed behavior and magnitudes of $\varepsilon_{Ram}$ and $\varepsilon_{Int}$ so different? What is the nature of the non-linearity in $\varepsilon_{Int}$ and can we quantify it?

**The relation between stress-strain curves and crumpling**

We believe the disparity between $\varepsilon_{Ram}(\sigma)$ and $\varepsilon_{Int}(\sigma)$ is a signature of crumpling and can be understood by clarifying the definition of strain. The shifts of Raman peaks, and hence $\varepsilon_{Ram}(\sigma)$ derived from them, reflect length changes of the carbon-carbon (C-C) bonds. Quantitatively, $\varepsilon_{Ram} = (L - L_0)/L_0$, where $L_0$ and $L$ are the lengths of the membrane before and after the application of stress. The "true" length of the membrane $L$ is not affected by crumpling provided C-C bond lengths are unchanged [45]. On the other hand, interferometric profilometry senses the profile of the membrane averaged with micrometer resolution, $\varepsilon_{Int} = (L^{AV} - L_0^{AV})/L_0^{AV}$, where $L_0^{AV}$ and $L^{AV}$ are the lengths of the *averaged* profiles. Thus defined $L^{AV}$ decreases when the membrane is crumpled. The difference between $L$ (red lines) and $L^{AV}$ (dashed green lines) is illustrated in the cartoon of Fig. 3b showing cross sections of circular membranes under the application of stress. At zero applied stress, crumpling causes a large difference between the "true" length of the cross section, $L_0$, and the length of its averaged profile, $L_0^{AV}$. When the stress is large enough to suppress crumpling ($\sigma_*$), that difference vanishes and the true profile is virtually indistinguishable from the averaged profile, $L \sim L^{AV}$. Summarizing, $\varepsilon_{Ram}$ is the microscopic strain, relative change in the bond lengths or the change in true membrane length. Whereas $\varepsilon_{Int}$ is macroscopic strain, relative change in the length of the averaged profile.

This insight allows the following interpretation of the data. At small stress, the changes in $L^{AV}$ per unit stress are large compared to those in $L$ as the significant amount of "hidden" length contained in crumpling is being unraveled (Fig. 3b, middle). In the experimental data at $\sigma < \sigma_* \sim 1$ N/m, we indeed observe much larger $d\varepsilon_{Int}/d\sigma$ compared to $d\varepsilon_{Ram}/d\sigma$ (Fig. 3a). As the stress becomes larger, the amount of crumpling is gradually decreased. Finally, the crumpling is suppressed, the membrane is flat, and the difference between the change in $L$ and $L^{AV}$ disappears almost completely (Fig. 3b, right). Correspondingly, in standard devices at $\sigma > \sigma_* \sim 1$ N/m (Fig. 3a) or in strain engineered devices (Fig. 3a Inset, orange points) we observe $d\varepsilon_{Int}/d\sigma \sim d\varepsilon_{Ram}/d\sigma$ or equivalently $d\varepsilon_{Ram}/d\varepsilon_{Int} \sim 1$.

The near-constant difference $\Delta\varepsilon = \varepsilon_{Int}(\sigma) - \varepsilon_{Ram}(\sigma)$ observed in the regime of high stress is related to what is known as "hidden area" in geometry [22]. The hidden area $\Delta A_0$ is the difference between the true area of the membrane $A_0$ and the $A_0^{AV}$ area of its projection onto a plane parallel to the membrane at zero applied stress [46]. As evident from Fig. 3c, $\Delta A_0$ is the amount of area "hidden" in out-of-plane crumpling and is "revealed" when the membrane is stretched. From simple geometrical considerations, $\Delta\varepsilon \approx (L_0^{AV} - L_0)/L_0 \approx \frac{1}{2}(A_0^{AV} - A_0)/A_0 = \frac{1}{2}\Delta A_0/A_0$. We use the relative hidden area $\Delta A_0/A_0$ extracted from $\Delta\varepsilon$ to quantify the amount of crumpling in our devices. We obtain relatively large $\Delta A_0/A_0$ of 0.6, 0.8 and 1.0 % for devices A, B, and C respectively.

**Exploring the nonlinear response**

Having obtained a quantitative measure for crumpling strength, we further investigate the non-linear behavior of the macroscopic strain ($\varepsilon_{Int}$) relevant for most experiments. Recently, a theory [47] was

developed to describe the "anomalous Hooke's law" in the stress-strain relationship of crumpled graphene:

$$\varepsilon(\sigma) = \frac{\sigma_*}{\tilde{E}_{2D}}\left[\frac{\sigma}{\sigma_*} + \frac{1}{\alpha}\left(\frac{\sigma}{\sigma_*}\right)^\alpha\right] \quad (1)$$

Here, $\alpha$ is an exponent which determines the degree of non-linearity caused by crumpling and $\sigma_*$ is the "crossover stress", a measure of the stress required to flatten the membrane. Qualitatively, the mechanical behavior described by Eq. 1 is that of two springs in series. The first linear "spring", with stiffness $\tilde{E}_{2D} \sim 400$ N/m describes stretching of C-C bonds, while the second, non-linear "spring" corresponds to uncrumpling of a membrane. The theory of Ref. 47 predicts $\alpha \sim 0.1$ for static disorder (wrinkling) and $\alpha \sim 0.5$ for thermal fluctuations (flexural phonons).

The comparison of our experimental data with the predictions of Eq. 1 is greatly facilitated by our complementary measurements of $\varepsilon_{Int}$ and $\varepsilon_{Ram}$. By taking the difference $\varepsilon_{Int}(\sigma) - \varepsilon_{Ram}(\sigma)$, we isolate the contribution of the nonlinear term in Eq. 1 pertaining to the mechanics of crumpling. To account for built-in stress in our devices, we subtract an additional term $\varepsilon_0 = \varepsilon(\sigma_0)$ from Eq. 1, where $\sigma_0$ is built-in stress. This allows us to compare our data (where only applied strain is measured) with Eq. 1. We are then able to fit our experimental data for devices A, B, and C to the non-linear component in Eq. 1 with $\tilde{E}_{2D}$ determined from interferometry at high stress and $\alpha, \sigma_*,$ and $\sigma_0$ treated as free parameters.

Figure 4a illustrates the adherence of our data to the non-linear model. For all standard devices, we retrieve an average exponent $\alpha = 0.12 \pm 0.02$. This is close to $\alpha = 0.1$ expected for statically wrinkled graphene, confirming our earlier interpretation that static wrinkling rather than flexural phonons is the primary contributor to crumpling [29]. The average value of built-in stress obtained from the fit, $\sigma_0 = 0.07 \pm 0.01$ N/m, is close to what is observed by others [42,48]. The average cross-over stress was found to be $\sigma_* = 0.8 \pm 0.1$ N/m. Physically, this means a stress of at least 0.8 N/m was required to flatten the sample and retrieve a linear response at higher stress. In agreement with that, linear $\varepsilon(\sigma)$ was observed for the strain-engineered device where we estimate $\sigma_0 = 0.84 \pm 0.02$ N/m ($> \sigma_*$). It should be noted that the fits are not perfect indicating that there are facets of our experimental data not accounted for by the model. Possible reasons for deviations include: non-uniform stress fields, non-random wrinkle distribution, deviation of the geometry from perfectly circular, and possible presence of contaminants [49,50].

The notion of the hidden area can be further used to compare the data to prediction of the model of Ref. 47. There, the degree of crumpling was controlled by the "disorder parameter": $B \propto (\sigma_* - \sigma_0)/\tilde{E}_{2D}$. In Fig. 4b, parameter $B$ extracted from our fits vs. $\Delta A_0/A_0$ is plotted. The correlation seen in Fig. 4b means that higher crumpling measured experimentally does, in fact, correspond to higher disorder in the model.

**Conclusion**

In conclusion, we observed the crossover from nonlinear mechanical response of graphene in the regime of low applied stress to linear response at high stress. The degree of nonlinearity and the crossover stress were found to depend on the amount of crumpling. We determined the latter, as quantified by the hidden area, through complementary Raman spectroscopy and interferometry measurements. Our data is in good agreement with recent theoretical predictions of the "anomalous Hooke's law" in crumpled membranes. Furthermore, we have demonstrated the distinction between experimentally measuring the microscopic or macroscopic mechanical response of materials.

We would like to highlight a few possible applications of our results. First, in many nanomechanics experiments, the linear mechanical response of graphene and other 2D materials is assumed in the regime of low stress (e.g. Refs [51,52]). The conclusions of some of these works may need to be reassessed. Second, our results suggest that the mechanical constants of graphene can be engineered in a wide range by tailoring the amount of crumpling through strain engineering. Extremely soft devices may be useful, for example, as exquisite force sensors. Finally, the most exciting area for future work is at the intersection between condensed matter and statistical physics where it may be possible to study renormalization of elastic constants of crystalline membranes due to flexural phonons [53,54] and the competition between static and dynamic sources of disorder [16].

**Acknowledgements**

We acknowledge enlightening conversations with I. V. Gornyi, V. Y. Kachorovskii, and A. D. Mirlin as well as financial support from Defense Threat Reduction Agency Basic Research Award # HDTRA1-15-1-0036, NSF CAREER 4-20-632-3391, and the Sloan Foundation. The optical profilometry measurements and fabrication of the chips for graphene transfer were conducted at the Center for Nanophase Materials Sciences, which is a DOE Office of Science User Facility.

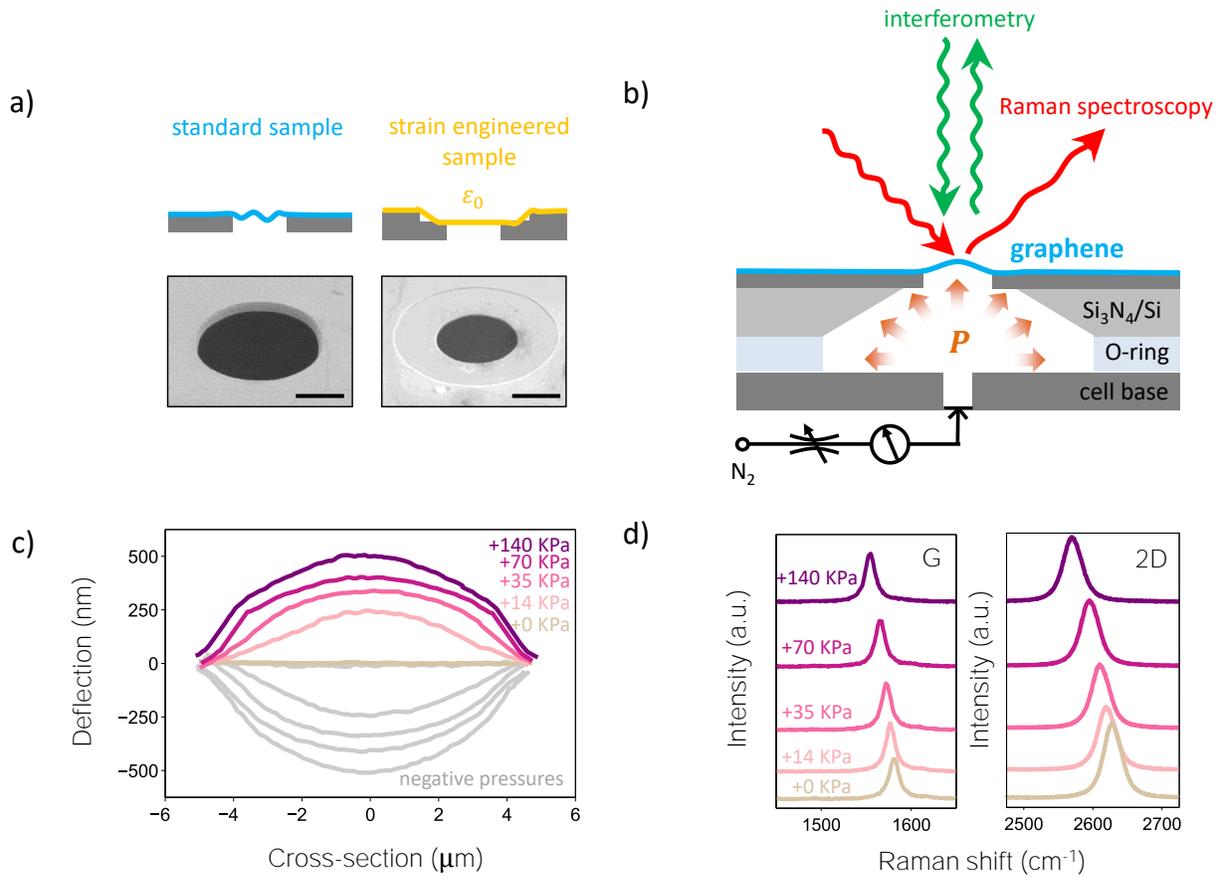

**FIG 1: Experimental set-up. a)** Top row: Cartoon views of standard and strain-engineered devices. Bottom row: scanning electron microscopy (SEM) images of representative samples (scale bar is 5 μm). **b)** Device schematic showing the application of pressure and our two measurement techniques: interferometry and Raman spectroscopy. Depending on the orientation of the sample chip we can apply positive (away from the sample, as pictured) or negative (towards the sample) pressures. **c)** Membrane profiles for both positive and negative pressures as measured by wide-field interferometry. **d)** Raman spectra of graphene showing the G and 2D Raman peaks throughout the range of applied pressure.

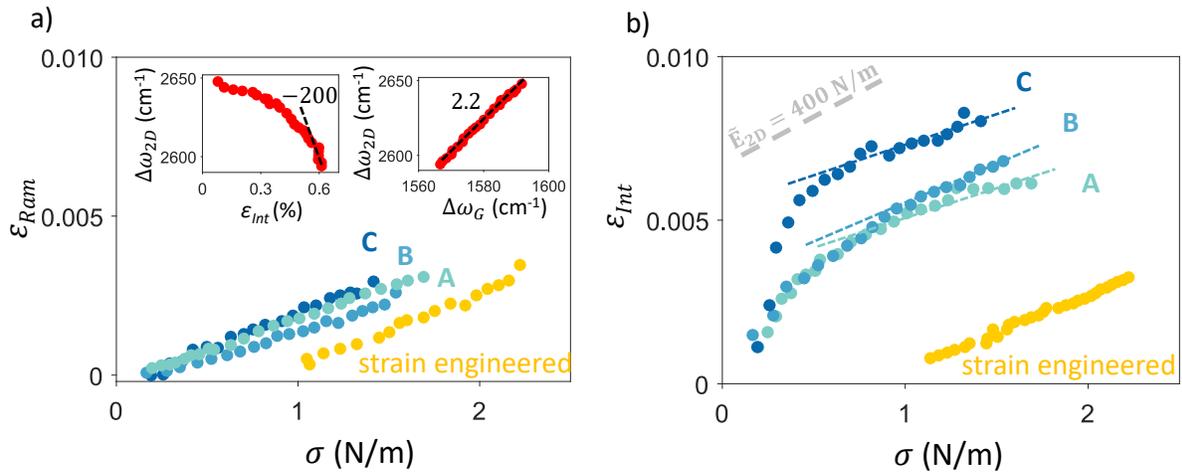

**FIG 2: Stress-strain curves from interferometry and Raman spectroscopy. a)** Stress-strain as determined from Raman spectroscopy, $\varepsilon_{Ram}(\sigma)$, for three standard samples A, B, and C (blue points) along with a strain engineered device (orange points). The data for the strain-engineered device is offset for clarity. Left inset: The progression of Raman 2D peak shift vs. $\varepsilon_{Int}$ used to calibrate peak sensitivity $\partial\omega/\partial\varepsilon$ (dashed black line). Right inset: The position of the 2D Raman peak plotted vs. the position of the G Raman peak. The slope of 2.2 indicates that changes in peak positions are due to strain. **b)** Stress-strain as determined from interferometry, $\varepsilon_{Int}(\sigma)$, for the same devices shown in a). Dashed grey line shows slope expected for flat graphene with the stiffness $\tilde{E}_{2D} = 400$ N/m. Dashed colored lines indicate the region of linear mechanical behavior.

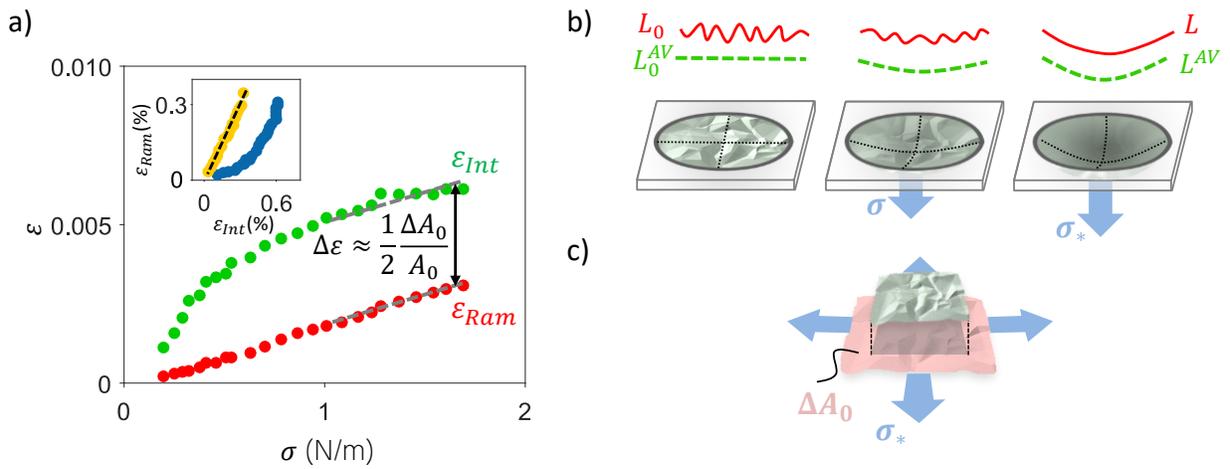

**FIG 3: The relation between strain and crumpling. a)** The comparison of the strain measured via interferometry ($\varepsilon_{Int}$, green curve) and the strain determined via Raman spectroscopy ($\varepsilon_{Ram}$, red curve) vs. applied stress $\sigma$ for device A. Inset: $\varepsilon_{Ram}$ vs $\varepsilon_{Int}$ for the same device shown in the main panel (blue points) and strain engineered device (orange points). Dashed black line has slope ~1. **b)** Cartoon illustrating the evolution of crumpling in a membrane under gradually increasing stress. Cross-section of the membrane and the same cross-section averaged with micron resolution are shown above each membrane. **c)** Visualization of hidden area $\Delta A_0$ of a membrane.

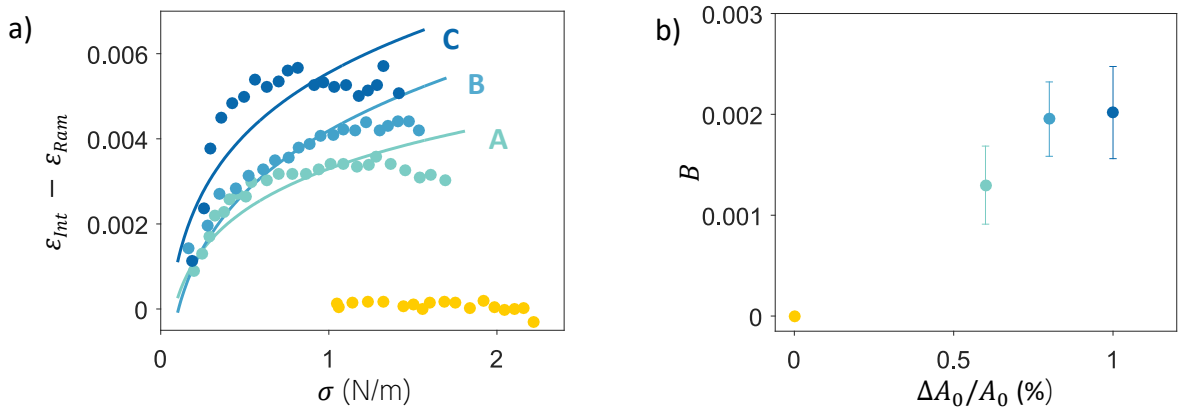

**FIG 4: Nonlinear mechanics in crumpled graphene. a)** The difference between the strain extracted from interferometry and the strain from Raman $\varepsilon_{Int} - \varepsilon_{Ram}$ vs. stress $\sigma$ for standard samples A, B, C (blue points) and the strain-engineered device (orange points). Solid lines are fits to the non-linear model described in the main text ($\varepsilon \propto \sigma^\alpha$). **b)** Disorder parameter $B$ vs. hidden area $\Delta A_0/A_0$.

# Supplementary Information for:
# Hidden area and mechanical nonlinearities in freestanding graphene

Ryan J.T. Nicholl[1], Nickolay V. Lavrik[2], Ivan Vlassiouk[3], Bernadeta R. Srijanto[2], and Kirill I. Bolotin[1,4]

[1]Department of Physics and Astronomy, Vanderbilt University, Nashville, TN 37235, USA
[2]Center for Nanophase Materials Sciences, Oak Ridge National Laboratory, Oak Ridge, TN 37831, USA
[3]Energy & Transportation Science Division, Oak Ridge National Laboratory, Oak Ridge, TN 37831, USA
[4]Department of Physics, Freie Universität Berlin, Arnimallee 14, Berlin 14195, Germany

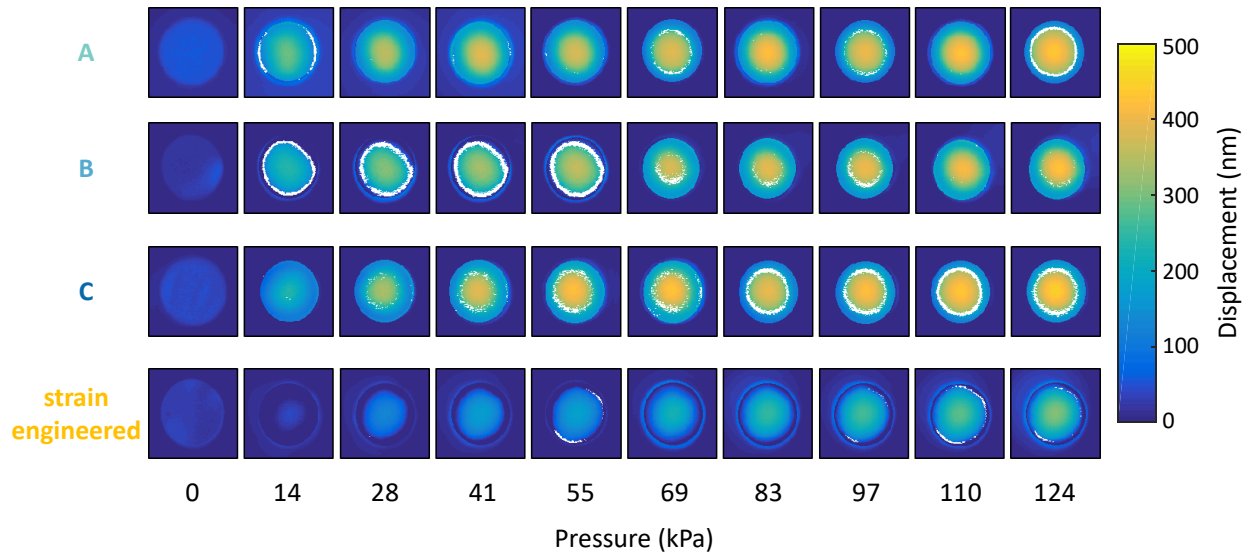

**Supplementary Fig 1:** Full interferometric data for standard samples A, B, C, and a strain engineered sample at different pressures. Color is out-of-plane displacement. Missing data are labelled white. Center point deflection $h$ was extracted from this data by taking the height difference between the center and the edge of the membrane. The result was cross-checked for consistency by determining the radius of curvature $R$ of the membrane and calculating the deflection as $h = \frac{a^2}{2R}$. We note that the center point deflection is unaffected by overall height of the sample.

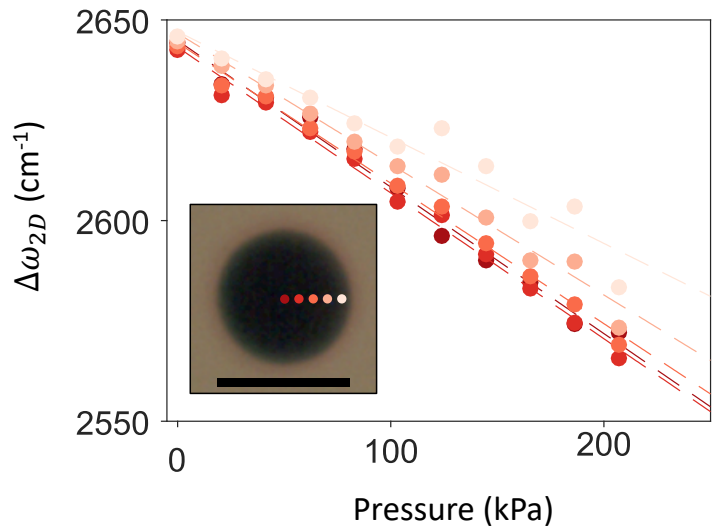

**Supplementary Fig 2:** Dependence of Raman spectra on spot position. Raman spectra were taken in 5 locations at 1 µm increments away from the center of a pressurized, non-strain-engineered device These positions are indicated on the photograph of the device (inset, scalebar is 10 µm). The main panel shows the resulting shifts of the 2D Raman peak vs. pressure, the dashed lines are linear fits to the data. The spectra show identical trends within ~2 µm of the membrane center – the slope varies no more than ~4 %. Further away from the center, the response changes. For example, 4 µm away from the center there is a ~30 % reduction in the magnitude of slope. This is expected: in pressurized bulge test set-up, only the center of the membrane is under perfect biaxial strain. The Raman spectra we report in the main text are at the center of the membrane, with accuracy better than ~2 µm.

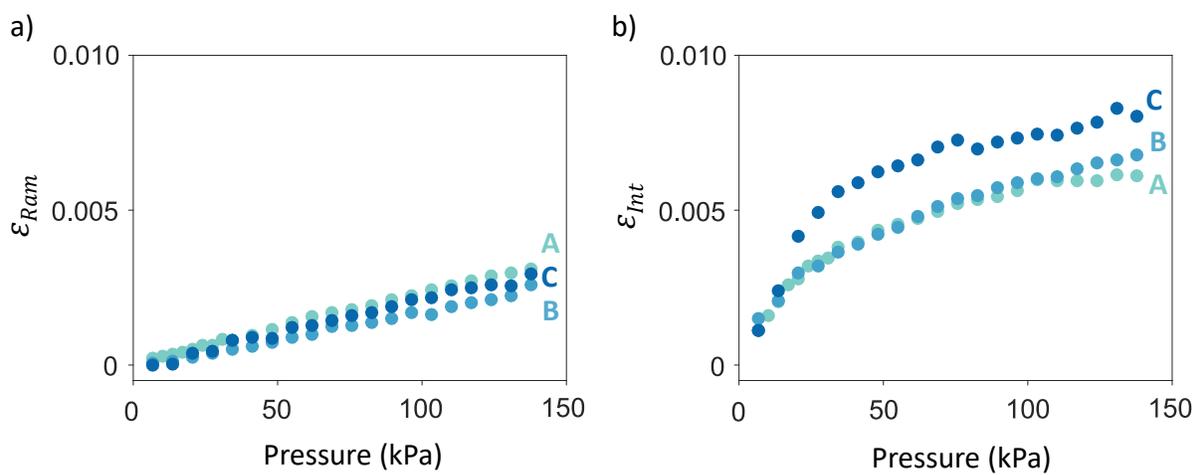

**Supplementary Fig 3:** Strain determined from Raman spectroscopy (a) and strain determined from interferometry (b) vs. applied pressure $P$. We see that the trends are qualitatively similar to the results in the main text – $\varepsilon_{Ram}(P)$ is linear like $\varepsilon_{Ram}(\sigma)$ and $\varepsilon_{Int}(P)$ is nonlinear like $\varepsilon_{Int}(\sigma)$.

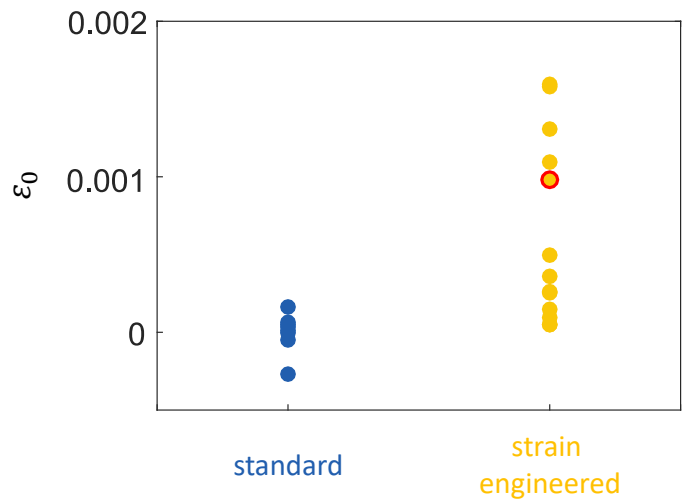

**Supplementary Fig 4:** Built-in strain of standard and strain engineered samples. The strain is determined at $P = 0$ from Raman spectroscopy as described in the main text assuming $\partial \omega^{2D}/\partial \varepsilon = -150$ cm$^{-1}$/%, $\omega_0^{2D} = 2650$ cm$^{-1}$, $\partial \omega^{G}/\partial \varepsilon = -75$ cm$^{-1}$/%, and $\omega_0^{G} = 1580$ cm$^{-1}$. To account for possible doping effects, we then define $\varepsilon_0 = 0$ as the average of the standard devices. Red circled device is the strain engineered device studied in the main text. Evidently, the built-in stress varies from device to device. This is due to multiple factors: slipping of graphene during transfer, built-in strain in pre-grown graphene, and the character of wrinkling.